\documentclass[12pt]{article}
\usepackage{amsmath}
\usepackage{amscd}
\usepackage{amssymb}
\usepackage{array}

\begin{document}
\rightline{CERN-TH/2002-030}

\newcommand{\R}{\mathbb{R}}
\newcommand{\C}{\mathbb{C}}
\newcommand{\Z}{\mathbb{Z}}
\newcommand{\Hb}{\mathbb{H}}
\newcommand{\id}{\relax{\rm 1\kern-.35em 1}}

\newcommand{\rUSp}{\mathrm{USp}}
\newcommand{\rO}{\mathrm{O}}
\newcommand{\rSO}{\mathrm{SO}}
\newcommand{\rSU}{\mathrm{SU}}
\newcommand{\rU}{\mathrm{U}}
\newcommand{\rSL}{\mathrm{SL}}
\newcommand{\rE}{\mathrm{E}}

\newcommand{\fsl}{\mathfrak{sl}}
\newcommand{\fso}{\mathfrak{so}}
\newcommand{\ft}{\mathfrak{t}}

\newcommand{\be}{\mathbf{E}}

\vskip 1.5cm

  \centerline{\LARGE \bf On the  Super Higgs Effect  }

  \bigskip

   \centerline{\LARGE \bf in Extended  Supergravity }

 \vskip 3cm
\centerline{L. Andrianopoli$^\sharp$, R. D'Auria$^\sharp$,  S.
Ferrara$^\flat$ and M. A. Lled\'o$^\sharp$.}

\vskip 1.5cm

\centerline{\it $^\sharp$ Dipartimento di Fisica, Politecnico di
Torino,} \centerline{\it Corso Duca degli Abruzzi 24, I-10129
Torino, Italy and } \centerline{\it   INFN, Sezione di Torino,
Italy. }

\medskip

\centerline{\it $^\flat$ CERN, Theory Division, CH 1211 Geneva 23,
Switzerland, and } \centerline{\it INFN, Laboratori Nazionali di
Frascati, Italy}

\vskip 1cm

\begin{abstract}
We consider the reduction of supersymmetry in $N$-extended four
dimensional supergravity via the super Higgs mechanism in theories
without cosmological constant.
 We provide an analysis largely based on the properties of long and short multiplets of Poincar\'e supersymmetry.
 Examples of the super Higgs phenomenon are realized in spontaneously broken $N=8$ supergravity through the
  Scherk-Schwarz mechanism and in superstring compactification in presence of brane fluxes. In many models the
   massive vectors count the difference in number of the translation isometries of the scalar $\sigma$-model
    geometries in the broken and unbroken phase.
\end{abstract}

\vfill\eject

\section{Introduction}

If supersymmetry is of any relevance in Nature it must be realized in a broken phase.
 Since supersymmetry extended to
 curved spacetime becomes a gauge theory, called supergravity, supersymmetry breaking must
  be spontaneous and therefore the
  super Higgs mechanism must take place. It is then of physical interest to study  the spontaneous
  breaking from  $N$ to $N'$
supersymmetries. In particular, the breaking $N\rightarrow 1\rightarrow 0$ is relevant for the hierarchy
 problem if supersymmetry has to solve it.

In the present paper we analyze some general features of
supergravity theories in dimension $d=4$ with scalar potentials \cite{dwn,war} allowing
 flat
Minkowski background\footnote{ We do not consider here super Higgs
effect in curved (AdS) backgrounds, which has been studied in many
examples in the literature  \cite{war,hw1}}.
 Given an unbroken theory with $N$ supersymmetries, we analyze when the degrees of freedom are consistent
  with the existence of a broken phase, which retains $N'$ supersymmetries.
Our analysis,  based on properties of massless and massive
representations \cite{fsz,st} of Poincar\'e supersymmetry,
 is mostly kinematic. More constraints are expected to come from the dynamical realization of the spontaneously
  broken theory.

For $N\rightarrow N'\geq 3$ \cite{ccfdfm} the analysis is particularly predictive, since the massless  theory
 is completely fixed by supersymmetry. This means that, assuming that there is a phase transition between the
  unbroken and the broken theories, the result of integrating out  the massive modes must give as a result the
   only theory allowed by  $N'$ supersymmetry.

An important difference in the super Higgs mechanism occurs depending whether $N-N'$ is even or odd. This is
 because spin $3/2$ BPS (short)  massive  multiplets can only occur in pairs since they carry a (BPS) charge,
  thus needing two of them to  form a  CPT invariant multiplet. So when $N-N'$ is odd at least one multiplet
  must be long (since in this case  it is  CPT invariant by itself). We will see that this condition already
  excludes some possibilities.

When $N'\leq 4$ there are  massless matter multiplets ($\lambda_{MAX}\leq 1$) of the reduced $N'$ supersymmetry.
 They  can then undergo a Higgs mechanism and become massive. So, in abscence of further dynamical informations,
  one can predict only the maximal number of residual massless multiplets.

A general pattern emerges by studying the supergravity models which admit a spontaneously broken phase with
 Minkowski background (vanishing cosmological constant). The  isometry  group $G$ has an abelian subalgebra
  that acts as translations on $G/H$. The broken  gauge symmetries  belong to this subalgebra \cite{cgp,fgp,tz}.
   This is at least true both, in the spontaneous breaking through Scherk-Schwarz \cite{ss,css} mechanism and in
    the breaking $N=4\rightarrow N=3$
 \cite{fp,kst} through compactification in the presence of  brane fluxes \cite{ps,tv,ma,cklt,gkp}. A detailed
 study of the pattern of symmetry breaking and its relation to scalar geometry will be considered elsewhere.
The construction of four dimensional gauged supergravity can be
found in the literature \cite{hw1},\cite{dwn},\cite{hu},\cite{hw}. A
partial classification  was given in \cite{cfgtt}.

The generalized dimensional reduction of  Scherk-Schwarz  has been
shown to have a pure four dimensional interpretation as a gauged
N=8 supergravity \cite{adfl} with a ``flat group" (in the notation
of Scherk and Schwarz) as gauge group. Our discussion of the super
Higgs effect in supergravity  is limited to those
 superstring models or higher dimensional theories in which the spontaneous breakdown does not involve neither
 stringy nor Kaluza-Klein modes. In this situation,  one may hope that the discussion of spontaneous
  supersymmetry breaking can be
 confined to an effective field theory which involves only a small number of degrees of freedom,
  both for the massless and the massive sectors. 
This has been shown to occur in K-K supergravities where, as an example, the de Wit -- Nicolai gauged $SO(8)$ theory
is a consistent truncation of M-theory on $AdS_4 \times S^7$ \cite{dwn2}.
This is because the masses of the effective models
  that we consider here
  can be taken much smaller than the Kaluza-Klein masses or the string scale \cite{fp}.  \footnote{Examples
 where this is not possible have been considered in the literature, when keeping only  massless states \cite{dlp,bkl}
 or when keeping also
  the massive ones \cite{kk}.}.

The paper is organized as follows. In Section 2. we recall  the structure of massless and massive
(long and short) Poincar\'e  supermultiplets in four dimensions. In particular, we treat in detail
the massive multiplets with maximum spin 3/2 which are relevant for the super Higgs effect.

In Section 3. we discuss spontaneously broken $N=8$ supergravity. We see that some patterns of
supersymmetry reduction  $N\rightarrow N'$ are not allowed in the super Higgs mechanism. Then we
consider all cases with $N=8,6\rightarrow 2\leq N'<N$.

In Section 4. we discuss the models which are dinamically realized through the Scherk-Schwarz
mechanism and infer the relation between the Higgs breaking and the broken symmetries of the scalar
geometry.

In Section 5. we discuss the relation of the mass generation of the vector bosons with the broken
 symmetries of the sigma model. We also  consider spontaneous sypersymmetry breaking in matter
  coupled theories which arise in string  compactifications in presence of brane fluxes.

In Section 6. we end with some concluding remarks.

\section{Massless and Massive Representations of extended supersymmetry in $d=4$}
We make here a short review of massless and massive representations  of the  $N$ extended super
 Poincar\'e algebra  in a space
time of dimension 4 and signature 2 \cite{fsz,st}. The odd part of the super Lie
algebra is arranged as a  direct sum of  $N$ Weyl (chiral) complex
spinor representations. A $\C$-linear basis of the odd generators
is
\begin{equation}\{Q^i_\alpha\}_{i=1,\dots N;
\; \alpha=1,2}.\label{basis1}\end{equation} We identify $\C^N$
with its dual by means of the sesquilinear form in $\C^N$,
$B(u,v)=u^\dagger v$. Then, the complex conjugates of
(\ref{basis1}) are denoted as
\begin{equation} (Q_\alpha^i)^*=\bar Q_{\dot\alpha
i}.\label{basis2}\end{equation}   (\ref{basis1}) and
(\ref{basis2}) span the odd part of the superalgebra over the real
numbers. The (anti) commutation relations are
 \begin{eqnarray}\{Q_\alpha^i,\bar Q_{\dot\alpha
 j}\}&=&2\sigma_{\alpha\dot\alpha}^\mu
 p_\mu\delta^i_j\nonumber \\\{Q_\alpha^i, Q_{\beta}^j\}&=&\epsilon_{\alpha\beta}Z^{ij}
 \nonumber\\\{\bar Q_{\dot\alpha i},\bar Q_{\dot\beta
 j}\}&=&\epsilon_{\dot\alpha\dot\beta}\bar Z_{ij}
 \label{susy}
 \end{eqnarray}
 where $p_\mu$ is the translation generator and $Z^{ij}$ are bosonic central generators
  organized in an
antisymmetric matrix.
 This is to be completed with the transformation of $Q_\alpha^i$ under the generators of
 the Lorentz group $M_{\mu\nu}$ as spinor representations and the
 commutation relations of the Poincar\'e generators among themselves.
  The automorphism group of the algebra is
$\rU(N)$, $Q$ and $\bar Q$ transforming in the fundamental
($\mathbf{N}$) and antifundamental ($\bar{\mathbf{N}}$)
representations of $\rU(N)$ respectively and $Z^{ij}$ in the two
fold antisymmetric representation. There is also a discrete
automorphism, the  CPT symmetry under which
 the generators of the algebra transform as
$$Q_\alpha^i\rightarrow i(Q_\alpha^i)^*=i\bar Q_{\dot\alpha i},\qquad Z^{ij}\rightarrow -(Z^{ij})^*=-\bar Z_{ij}.$$

The unitary representations of this superalgebra are obtained using the method of
 induced representations. One considers the orbit of the Lorentz group on the dual space to the translations.
  The orbits are given by the value of the invariant $p^2=m^2$. For $m>0$ the little group is SU(2) and for $m=0$
  it is  $E(2)$, from which we take representations which are non trivial only for the compact subgroup $U(1)$.
  The representation of the full Poincar\'e superalgebra is achieved by building at each point of the orbit a
   fiber which is a direct sum of representations of the little group, the odd generators mixing the different
   representation spaces.

We will take as conventions $\eta_{\mu\nu}=\mathrm{diag}(-1,+1,+1,+1)$ and
$$\sigma_0=\begin{pmatrix}1&0\\0&1\end{pmatrix},\quad \sigma_1=\begin{pmatrix}0&1\\1&0\end{pmatrix}
   \quad \sigma_2=\begin{pmatrix}0&-i\\i&0\end{pmatrix},\quad
   \sigma_3=\begin{pmatrix}1&0\\0&-1\end{pmatrix}.$$

\paragraph{Massless representations}
 To see what is the fiber at one point of the orbit we take the point  $p^\mu=(E,0,0,-E)$ (rest frame).
  The central charges must
be set to zero in order to obtain a unitary representation.  The
algebra (\ref{susy}) becomes

\begin{eqnarray}\{Q^i,\bar
Q_{j}\}&=&2\begin{pmatrix}0&0\\0&2E\end{pmatrix}
 \delta^i_j\nonumber \\\{Q_\alpha^i, Q_{\beta}^j\}&=&0
 \label{susymassless}
 \end{eqnarray}
 We see that the generators $Q^i_1,\;\bar Q_{1j}$ form an abelian superalgebra that decouples from the
 rest. The nontrivial anticommutation relations  are of  $N$
 creation and $N$ annihilation  operators. So the representations
 are constructed by giving a vacuum state $|\Omega\rangle_\lambda$
 $$Q_2^i|\Omega\rangle_\lambda=0$$ with
 helicity (representation of the little group)  $\lambda $
and acting on it with the creation operators
 $\bar Q_{2i}$.
 $$|i_1,\dots i_k\rangle_\lambda=\frac{1}{k!(2\sqrt E)^k}\bar
 Q_{2i_k}\cdots \bar Q_{2i_1}|\Omega\rangle_\lambda.$$
 Each creation operator lowers the helicity of the state by $1/2$.
 The state at level $k$ has helicity $\lambda-k/2$ and is in the
 $k$-fold
 antisymmetric representation of $\rU(N)$.
  The representation has  $2^N$ states. The helicities range from  $\lambda$ to $\lambda-N/2$.
  Since by CPT $\lambda \rightarrow -\lambda$, a CPT conjugate representation is obtained with a
  vacuum $|\Omega\rangle_{N/2-\lambda}$. The direct sum is CPT invariant and has dimension $2^{N+1}$.
   Notice that CPT  also changes the k-fold antisymmetric representation of $\rU(N)$, $[k]$ by its
    complex conjugate $[\bar k]\approx [N-k]$. There is one case where the CPT doubling is not required.
     This happens when  $\lambda=N/2-\lambda$ (so  $N$ is  necessarily even) and the representation
      $[N/2]$ (corresponding to the spin 0 state) is real. The last condition requires $N$ to be a multiple of 4.

We can also consider a vacuum which is a non trivial representation $R$ of the automorphism group $\rU(N)$.
 The vacuum will then be labelled by $|\Omega\rangle_{\lambda,R}$, and the helicity states will be in the
  tensor product representation $[k]\otimes R$. The CPT conjugate representation will then be
  $|\Omega\rangle_{N/2-\lambda,\bar R}$.

>From the above it follows that in massless multiplets the helicity range is $|\Delta \lambda|=N/2$ so
 that if the maximum helicity $\lambda _{MAX}$ of the multiplet is $|\lambda _{MAX}|\leq 2$ then
  necessarily $N\leq 8$, if $|\lambda _{MAX}|\leq 1$ then  $N\leq 4$ and if $|\lambda _{MAX}|\leq 1/2$ then
  $N\leq 2$.

We report in Tables \ref{hel23/2} and \ref{hel11/2} the massless representations for $N\leq 8$
 \footnote{We don't consider $N=7$ because $N=7$ supergravity is the same as $N=8$ supergravity.}
  with  $\lambda_{MAX}\leq 2$.  A couple of  states of helicity $\pm\lambda$ are denoted as  $(\lambda)$,
   the number in front is their multiplicity. Before the doubling, the multiplicity is   the dimension  of
    the representation $[k]$ of $\rU(N)$, $\begin{pmatrix}N\\k\end{pmatrix}$. After the doubling the multiplicity
    is the dimension of the representation $[k]\oplus[4\lambda-k]$.

\paragraph{Massive representations}  In the rest frame we have
that the  momentum is $p_\mu=(M,0,0,0)$.  $Q$ and $\epsilon\bar Q$
transform the same
 representation of the little group SU(2), so we can define
 $Q^a_\alpha, \quad a=1,\dots 2N$
$$Q^a_\alpha=Q_\alpha^i,\; a=i=1,\dots N;\quad
Q_\alpha^a=\epsilon^{\dot \alpha\dot\beta} Q^*_{\beta i},\;
a=N+i=N+1,\dots 2N.$$  This definition can be understood as a
reality condition on general complex vectors  $Q_\alpha^a$. This
condition is preserved by a transformation of the group USp($2N$).
The operators $Q_\alpha^a$ at the selected point of the orbit must
satisfy the relations
$$\{Q_\alpha^a,Q_\beta^b\}=\epsilon_{\alpha\beta}\Lambda^{ab},
\qquad
 \Lambda=\begin{pmatrix}Z&M\id\\-M\id&Z^*\end{pmatrix}.$$
$\Lambda$ is an antisymmetric  quaternionic matrix, that is,
$\Lambda^*=-\Omega\Lambda\Omega$ with $$\Omega=\begin{pmatrix}
0&\id\\-\id&0\end{pmatrix}.$$ The quaternionic property is
preserved by an USp($2N$) transformation, which is an automorphism
of the algebra in the rest frame that commutes with SU(2).  In
particular, this means
 that with a transformation $U\in \rUSp(2N)$ we can bring $\Lambda$ to a skew diagonal form
 $\Lambda'=U\Lambda U^T$ \cite{dfl}
$$\Lambda'=\begin{pmatrix}0_{n\times n}&\rho_{n\times n}\\
-\rho_{n\times n}&0_{n\times n}\end{pmatrix}$$\begin{equation}\rho_{n\times n}=
\begin{pmatrix}M\id_{2\times 2}+z_1\sigma_3&0\cdots&0\\
0&M\id_{2\times 2}+z_2\sigma_3&\cdots 0\\ \cdots\\
0&0&\cdots&M\id_{2\times 2}+z_n\sigma_3
\end{pmatrix},\label{lambda}\end{equation}
with $n=N/2$ in the even case and $n=(N-1)/2$ in the odd case. $U$ can be interpreted as a change
 of basis of the $Q$'s
\begin{equation}\{Q_\alpha,Q_\beta\}=\epsilon\Lambda'.\label{anticomassive}
\end{equation}
From this it can be seen that unitary representations are obtained
only if $M\geq |z_i|$ (BPS bound).

\subparagraph{No central charges} The cases   $z_i=0$ or $z_i\neq
0$ but  $M>|z_i|$ are qualitatively the same.
 From (\ref{lambda}) and (\ref{anticomassive}) we see that we can make a rescaling of the $Q$'s and
  we have an algebra with $2N$ creation and $2N$ annihilation operators. It shows explicit invariance
  under $\rSU(2)\times\rUSp(2N)$. The vacuum state is now labeled by the spin representation of SU(2),
   $|\Omega\rangle_J$. If $J=0$ we have the fundamental massive multiplet with $2^{2N}$ states. These
   are organized in  representations of SU(2) with $J_{MAX}=N/2$. With respect to $\rUSp(2N)$ the states
   with fixed $0<J<N/2$ are arranged in the $(N-2J)$-fold $\Omega$-traceless antisymmetric representation,
    $[N-2J]$.

The general multiplet with a spin $J$  vacuum can  be obtained by tensoring the fundamental multiplet
 with spin $J$ representation of SU(2). The total number of states is then $(2J+1)\cdot2^{2N}$.

Massive multiplets with $z_i=0$ are called long multiplets or non BPS states. The only difference with
multiplets $z_i\neq 0$ and $M>|z_i|$ is that the last ones must be doubled in order to have PCT invariance,
since $z_i\rightarrow -z_i$ under PCT. We will no longer  consider this case.

\subparagraph{BPS multiplets} If $q$ of the eigenvalues $z_i$
saturate the BPS bound, $M=|z_i|$, $i=1,\dots q$ then $q$, of the
pairs creation-annihilation operators have abelian anticommutation
relations and are totally decoupled, similarly to the phenomenon
occuring for $m=0$. The resulting multiplets are said to be $q/N$
BPS. Note that $q_{MAX}=N/2$ for $N$ even and $q_{MAX}=(N-1)/2$
for $N$ odd. The $\rUSp(2N)$ symmetry is now reduced to
$\rUSp(2(N-q))$.

The reduced or short multiplet has the same number of states than a long multiplet of the $N-q$
supersymmetry algebra. The fundamental multiplet, with $J=0$ vacuum  contains $2\cdot2^{2(N-q)}$
 states with $J_{MAX}=(N-q)/2$. Note the doubling due to CPT invariance.
 Generic massive short multiplets can be obtained by making the tensor product
 with a spin $J_0$ representation of SU(2).

\bigskip

In the discussion of the super Higgs and Higgs  effect, massive multiplets with spin 3/2 and 1 are
 relevant. In Tables \ref{spin3/2}, \ref{spin1} and \ref{spin1/2} we give a list of all possible
  cases for $N\leq 8$. The occurence of long spin 3/2 multiplets is only possible for $N=3,2$ and long spin
   1 multiplets for $N=2$. In $N=1$ there is only one type of massive multiplet (long) since there are
    no central charges. Its structure is
$$\bigl[(J_0+\frac{1}{2}),2(J_0),(J_0-\frac{1}{2})\bigr],$$ except
for $J_0=0$ that we have
 $\bigl[(\frac{1}{2}),2(0)\bigr]$.

In the tables we will denote the spin states by $(J)$ and the
number in front of them is their multiplicity. In the fundamental
multiplet, with spin $J_0=0$ vacuum, the multiplicity of the spin
$(N-k)/2$  is the dimension of the  k-fold antisymmetric
$\Omega$-traceless  representation of $\rUSp(N)$. For multiplets
with $J_0\neq 0$ one has to make the tensor product of the
fundamental multiplet with the representation of spin $J_0$. We
also indicate if the multiplet is long or short.

\section{Super Higgs effect in supergravity: generalities}
When performing the  spontaneous breaking  of extended
supersymmetry from $N$ to $N'$  in supergravity, a necessary
requirement is that $N-N'$ of the original massless gravitinos
must describe massive representations of the unbroken $N'$
supersymmetries. Since massive spin 3/2 may occur both in long and
short multiplets only  for $N\leq 3$ ( Table \ref{spin3/2}),
strong constraints emerge if $N-N'$ is odd, because in this case
at least one spin 3/2 multiplets must be long. This immediately
excludes the cases $N=8,6\rightarrow N'=5$ and $N=5\rightarrow
N'=4$.

Other cases that are not possible are the ones described in the following.

\subparagraph{$N=6\rightarrow N'=3$.} We first write down the
decomposition of the massless, helicity 2, $N=6$ multiplet into
massless multiplets of $N=3$.
\begin{eqnarray*}\bigl[ (2), 6(\frac{3}{2}), 16 (1), 26(\frac{1}{2}), 30(0)\bigr]\rightarrow
\bigl[ (2), 3(\frac{3}{2}), 3(1), (\frac{1}{2})\bigr]+\\ 3\bigl[
(\frac{3}{2}), 3(1), 3(\frac{1}{2}), 2(0)\bigr]+4\bigl[ (1),
4(\frac{1}{2}), 6(0)\bigr].\end{eqnarray*} The multiplets with
$\lambda_{MAX}=3/2,1$ must be recombined into massive multiplets
of $N'=3$.
 From Table \ref{spin3/2} it is easy to see that there is no combination of the long and short
 multiplets with spin 3/2 that can match the number of gravitinos and vectors simultaneously. In fact,
 one short and one long spin 3/2 multiplets will need 14 vectors.

\subparagraph{$N=5\rightarrow N'=3$.}The decomposition into massless representations is
\begin{eqnarray*}\bigl[ (2), 5(\frac{3}{2}), 10 (1), 11(\frac{1}{2}), 10(0)\bigr]\rightarrow
\bigl[ (2), 3(\frac{3}{2}), 3(1), (\frac{1}{2})\bigr]+\\ 2\bigl[
(\frac{3}{2}), 3(1), 3(\frac{1}{2}), 2(0)\bigr]+\bigl[ (1),
4(\frac{1}{2}), 6(0)\bigr].\end{eqnarray*} We could try to combine
the last two ones into two short $N=3$ spin 3/2 multiplets,  but
the number of vectors is already bigger than the seven vectors at
our disposal.

\subparagraph{$N=5\rightarrow N'=2$.}The decomposition into massless representations is
\begin{eqnarray*}\bigl[ (2), 5(\frac{3}{2}), 10 (1), 11(\frac{1}{2}), 10(0)\bigr]\rightarrow
\bigl[ (2), 2(\frac{3}{2}), (1)\bigr]+\\ 3\bigl[ (\frac{3}{2}),
2(1), (\frac{1}{2})\bigr]+3\bigl[ (1), 2(\frac{1}{2}), 2(0)\bigr]+
\bigl[ 2(\frac{1}{2}), 4(0)\bigr].\end{eqnarray*} The minimal
combination of spin 3/2 multiplets that could be used is one long
and two short multiplets. But the number of states with  $J_z=0$
is already 12. They should match the number of helicity states in
the last three terms of the equation above, which is 10, so it is
impossible.

\subsection{Spontaneously broken  supergravity $N=8\rightarrow N'=6,4$
 \label{8to64}} We want to explore the $N=8\rightarrow N'=6,4$ spontaneous breaking of supersymmetry.
  In these cases there are no long spin 3/2 multiplets. In the models with $N'=4$ there
  appear multiplitets with $\lambda_{MAX}=1$, so there is a possibility of a further Higgs effect on these multiplets.

\subparagraph{$N=8\rightarrow N'=6$.}The decomposition into massless multiplets is as follows,
\begin{eqnarray*}\bigl[ (2), 8(\frac{3}{2}), 28 (1), 56(\frac{1}{2}), 70(0)\bigr]\rightarrow
\bigl[ (2), 6(\frac{3}{2}), 16(1),26(\frac{1}{2}), 30(0)\bigr]+\\
2\bigl[ (\frac{3}{2}), 6(1), 15(\frac{1}{2}),
20(0)\bigr].\end{eqnarray*} The two massless $\lambda_{MAX}=3/2$
multiplets can be reread as one massive, 1/2 BPS,  spin 3/2
multiplet of $N'=6$ (Table \ref{spin3/2}), so in principle the
Higgs effect is possible.

\subparagraph{$N=8\rightarrow N'=4$.}The decomposition into massless multiplets is as follows,
\begin{eqnarray*}\bigl[ (2), 8(\frac{3}{2}), 28 (1), 56(\frac{1}{2}), 70(0)\bigr]\rightarrow
\bigl[ (2), 4(\frac{3}{2}), 6(1),4(\frac{1}{2}), 2(0)\bigr]+\\
4\bigl[ (\frac{3}{2}), 4(1), 7(\frac{1}{2}), 8(0)\bigr]+6\bigl[
(1), 4(\frac{1}{2}), 6(0)\bigr].\end{eqnarray*} There are two
types of $N=4$ spin 3/2 massive multiplets, both of them short
with $q=1,2$. The number of massless vectors which are not in the
gravity multiplet  is (from above) 22. We have two possibilities

\noindent 1. Two $q=2$ spin 3/2 multiplets ((16 vectors) plus six massless vector multiplets.

\noindent 2. One $q=1$ and one $q=2$ spin 3/2 multiplets (20 vectors) plus two massless vector multiplets.

The rest of the states also matches. The massless vector
multiplets may undergo  a Higgs effect. Two massless vector
multiplets have the same number of states than one massive one. In
case 1. we  can have a Higgs effect from 6 to 4, 2 or 0 massless
vectors. In case 2.  we have a Higgs effect from 2 to 0.

\subsection{Spontaneously broken  supergravity  $N=6\rightarrow N'=4,2$\label{6to42}}

\subparagraph{$N=6\rightarrow N'=4$.}The decomposition into massless representations is
\begin{eqnarray*}\bigl[ (2), 6(\frac{3}{2}), 16(1), 26(\frac{1}{2}), 30(0)\bigr]\rightarrow
\bigl[ (2), 4(\frac{3}{2}), 6(1),4(\frac{1}{2}), 2(0)\bigr]+\\
2\bigl[ (\frac{3}{2}), 4(1), 7(\frac{1}{2}), 8(0)\bigr]+2\bigl[
(1), 4(\frac{1}{2}, 6(0)\bigr].\end{eqnarray*} The only
possibility is to take an $N=4$, $q=2$, spin 3/2 multiplet and two
massless vector multiplets.

\subparagraph{$N=6\rightarrow N'=2$.}The decomposition into massless representations is
\begin{eqnarray*}\bigl[ (2), 6(\frac{3}{2}), 16(1), 26(\frac{1}{2}), 30(0)\bigr]\rightarrow
\bigl[ (2), 2(\frac{3}{2}), (1)\bigr]+ 4\bigl[ (\frac{3}{2}),
2(1), (\frac{1}{2})\bigr]+\\7\bigl[ (1), 2(\frac{1}{2},
2(0)\bigr)+ 4\bigl[ 2(\frac{1}{2}), 4(0)\bigr].\end{eqnarray*} For
$N'=2$ we have long and short spin 3/2 multiplets. The only
possibilities allowed are:

\noindent 1. Two long multiplets and one short. Then the rest of
the states arrange into three massless  vector multipltes and one
hypermultiplet ($\lambda_{MAX}= 1/2$).

\noindent 2. Two short multiplets. The rest of the states combine into seven
massless vector multiplets and two hypermultiplets.

Again, the massless multiplets can be combined into massive ones,
this time with the possibility of   including a mass term for the
hypermultiplet.

\subsection{Spontaneously broken  supergravity  $N=8\rightarrow N'=3,2$\label{8to32}}
As it can be seen in Table \ref{spin3/2}, for $N'=2,3$ there are both, long and short spin 3/2 multiplets.

\subparagraph{$N=8\rightarrow N'=3$.} The decomposition into massless representations is
\begin{eqnarray*}\bigl[ (2), 8(\frac{3}{2}), 28(1), 56(\frac{1}{2}), 70(0)\bigr]\rightarrow
\bigl[ (2), 3(\frac{3}{2}), 3(1), (\frac{1}{2})\bigr]+\\ 5\bigl[
(\frac{3}{2}), 3(1), 3(\frac{1}{2}), 2(0)\bigr]+10\bigl[ (1),
4(\frac{1}{2}), 6(0)\bigr].\end{eqnarray*} The first possibility
that arises  with no more than  25 vectors is to take one long and
two short gravitino multiplets, with 22 vectors, so three
additional vector multiplets must be present. The number of states
with spin 1/2 and $J_z=0$ also match the states with  helicity
$\pm 1/2$ and 0 in the massless multiplets. One can further have a
Higgs effect that takes two massless  vector multiplets into a
massive one.

\subparagraph{$N=8\rightarrow N'=2$.} The decomposition into massless representations is
\begin{eqnarray*}\bigl[ (2), 8(\frac{3}{2}), 28(1), 56(\frac{1}{2}), 70(0)\bigr]\rightarrow
\bigl[ (2), 2(\frac{3}{2}), (1)\bigr]+ 6\bigl[(\frac{3}{2}), 2(1),
(\frac{1}{2})\bigr]\\+15\bigl[ (1), 2(\frac{1}{2}), 2(0)\bigr]+10
\bigl[ 2(\frac{1}{2}), 4(0)\bigr].\end{eqnarray*} For the first
time we note the occurrence of a multiplet with
$\lambda_{MAX}=1/2$ (hypermultiplet). We have 27 massless vectors.
We have   four possible combinations of spin 3/2 multiplets: six
long , four  long and one short, two short and two  long, and
three short. Let $2n$ be the number of long gravitino  multiplets;
then $6-2n$ is the number of short gravitino multiplets, $15-4n$
is the number of vector multiplets and $7-n$ is the number of
hypermultiplets. Later we will see further constraints on these
cases.

\section{Scherk-Schwarz mechanism for $N=8,6$ spontaneously broken supergravity\label{ssbreak}}

In the previous section we have stated some necessary conditions
for the super  Higgs effect to take place. Of more interest is to
know whether one can find dynamical models which have the  two
phases, with  broken and  unbroken symmetry. The Scherk-Schwarz
generalized dimensional reduction \cite{ss,css} provides many
examples of such systems.

The Scherk-Schwarz mechanism starts by considering $N=8$
supergravities and makes a generalized  dimensional reduction
ansatz to dimension 4, which generically depends on four
parameters $m_i,\; i=1,\dots 4$. We obtain a family of four
dimensional models, where the parameters are interpreted as
gravitino masses, which means that the supersymmetry has been
spontaneously broken. In fact, the gravitinos come in pairs of
equal mass. One  of the masses, say $m_1$ is set  to zero in order
to keep some supersymmetry \cite{fz}. If additionally  $m_2=m_3=0$
($m_4\neq 0$) then we have an $N'=6$ model; if $m_2$=0
($m_3,m_4\neq 0$) we obtain an $N'=4$ model and the $N'=2$ model
is obtained with only $m_1=0$. The masses acquired by the
different states of the graviton multiplet depend on the
parameters $m_i$. They are given in Table \ref{spectrum}.

The $N'=6$ model is unique (Section \ref{8to64}). The $N'=4$ model
corresponds to case 1. in Section \ref{8to64} with a further Higgs
mechanism in the vector sector. In fact, at generic values of
$m_3$ and $m_4$ the model has has 2   massless and 2   massive
vector multiplets and for $|m_3|=|m_4|$ we get 4 massless and 1
massive vector multiplet. The model with 6 massless vectors is not
realized in this context.

For $m_i\neq 0$ and $|m_i|\neq|m_j|, \; i=2,3,4$ we have
$N=8\rightarrow N'=2$ (Section \ref{8to32}).  In this model all
massive  spin 3/2 multiplets  are short. Then there are three
massless vector multiplets, 12 short massive vector multiplets
and 7 massive hypermultiplets. It corresponds to a Higgs version
of the $n=0$ model in Section \ref{8to32}.

If $|m_i|=|m_j|, \; i=2,3,4$ we have the maximal number of
massless vector multiplets, that is 9.

\bigskip

Similarly one could start with $N=6$ supergravity  (Section
\ref{6to42}). In this case the number of  parameters is three and
have the same interpretation as gravitino masses. The masses
acquired by the helicity states of the $N=6$ graviton multiplet
are given in Table \ref{spectrum6}.

For $N'=4$ the theory is completely fixed. The gravitinos are
$\frac{1}{2}$BPS  multiplets and we have two massless vector
multiplets.

For $N'=2$ the theory corresponds to the case labelled by 2. in
Section \ref{6to42}.  If $|m_2|=|m_3|$ there are 5 massless
vectors multiplets and if $|m_2|\neq|m_3|$  there are  3.

\section{Goldstone bosons and translational symmetries}

When the manifold parametrized by the scalars is a coset space
$G/H$, there is an abelian  algebra of isometries that is
contained in $G$. This algebra is the maximal abelian ideal of the
solvable algebra associated to the coset space via the Iwasawa
decomposition (see for example \cite{he}).  In the examples that
we analyze in this section we have two models with two coset
spaces, $G/H$ corresponding to the unbroken supersymmetry model
and $G'/H'$ corresponding to the model with partial breaking of
supersymmetry once  the massive modes have been integrated out.
We will denote by $\ft(G/H)$ and $\ft'(G'/H')$ the abelian
subalgebras associated to the respective cosets (here the
``$\ft$'' stands for translational). $t$ and $t'$ are respectively
the dimensions of these  subalgebras.

If $n_v$ and $n_v'$ denote the number of massless  vectors in each
theory, we find  in  all the models analyzed that $t-t'=n_v-n_v'$.
The solvable group obtained in the Iwasawa decomposition (now in
the group instead that in the algebra) is diffeomorphic as a
manifold to $G/H$.  This parametrization has been considered in
the literature to analyze U-dualities in string theory
\cite{adfft,lps,cjlp}. The generators of the maximal abelian ideal
act as  translations on $t$ of the coordinates of $G/H$, which
appear only through derivatives in the Lagrangian and
 which are flat directions of  the scalar potential. This suggests that, as a general
rule for a consistent Higgs effect, these particular coordinates
are the Goldstone bosons connected to the spontaneous breaking of
$\R^{n_v}$ to $\R^{n_v'}$, so they have been absorbed into the
vectors that have acquired mass.

We analyze first  examples that are obtained with the
Scherk-Schwarz mechanism. In  these cases one can prove that the
above considerations are actually valid \cite{ss,css,svn}. It
would be interesting to know the cases where this rule does not
hold.

\subparagraph{$N=8\rightarrow N'=6$.} The coset space of the
scalars in  $N=8$ supergravity is $\rE_{7,7}/\rSU(8)$ and the
dimension of the translational subalgebra is $t=27$ \cite{adfft}.
In $N'=6$ the coset is $\rSO^*(12)/\rU(6)$, and $t'=15$. So
$t-t'=12$. It is easy to see that $n_v-n_v'=28-16=12$.

\subparagraph {$N=6\rightarrow N'=4$} For $N'=4$ the coset is
$$\frac{\rSO(6,n)}{\rSO(6)\times\rSO(n)}\times
\frac{\rSU(1,1)}{\rU(1)}. $$$n$ is the number of vector multiplets
in the theory. We take $n=2$ (see Section \ref{6to42}). Then
$t-t'=15-7=8$. We have  that  $n_v-n_v'=16-8=8$.

\subparagraph{$N=8\rightarrow N'=2$.} For $N'=2$ we  have a
certain number of vector multiplets  ($n_1$) and hypermultiples
($n_2$) (see Section \ref{8to32}).

The minimal model($m_i\neq m_j,\; i,j= 2,3,4$ in the notation of
Section \ref{ssbreak}) corresponds to  $n_1=3$ (we take $n_2=0$),
and  the coset is
$$\frac{\rSU(1,1)}{\rU(1)}\times\frac{\rSU(1,1)}{\rU(1)}\times\frac{\rSU(1,1)}{\rU(1)}.
$$$t'=3$, so $t-t'=27-3=24$. We have that  $n_v-n_v'=28-4=24$.

The maximal model ($m_1= m_2=m_3$) corresponds to having $n_1=9$, (again we take $n_2=0$). The coset space is
$$\frac{\rSU(3,3)}{\rSU(3)\times\rSU(3)\times\rU(1)}.$$ In this case $t-t'=27-9=18$ and $n_v-n_v'=28-10=18$.

\subparagraph{$N=6\rightarrow N'=2$.} We consider  first the case
with three massless vectors in the $N'=2$ theory. Then
$t-t'=15-3=12$ and $n_v-n_v'=16-4=12$.

The case with 5 massless vectors ($m_2=m_3$) has coset

$$\frac{\rSU(1,1)}{\rU(1)}\times\frac{\rSO(2,4)}{\rSO(2)\times SO(4)}.$$
$t-t'=15-5=10$ and $n_v-n_v'=16-6=10$.

\subparagraph{$N=8\rightarrow N'=4$.} We take the case with two massless vector multiplets. Then the coset is
$$\frac{\rSO(6,2)}{\rSO(6)\times \rSO(2)}\times\frac{\rSU(1,1)}{\rSO(2)\times \rU(1)}.$$
We have $t-t'=27-7=20$ and $n_v-n_v'=28-8=20$.

\bigskip

The examples that follow can be obtained  in Type IIB superstring
 compactified on an orientifold $T^6/\Z_2$ in presence of brane fluxes  \cite{fp,kst} and in certain
  gauged supergravity theories \cite{tz}.
We  want to consider the spontaneous breaking of $N=4,3$
supergravities down to $N'=3,2$. In order to have a consistent
reduction it is necessary that  the scalar manifold of
 the broken theory is a submanifold of the unbroken one \cite{adf}. For the $N'=2$ case this is just
 an assumption since the effects of integrating out the massive modes could be more complicated.

\subparagraph{$N=4\rightarrow N'=3$.} We consider the $N=4$ model with six massless vector multiplets,

$$\frac{\rSO(6,6)}{\rSO(6)\times\rSO(6)}\times
\frac{\rSU(1,1)}{\rU(1)}, $$ with $t=15$. Note that the SU(1,1)
factor is not considered because SU(3,3)$\subset$SO(6,6) The
decomposition of the massless multiplets is

\begin{eqnarray*}\bigl[(2), 4(\frac{3}{2}), 6(1), 4(\frac{1}{2}),2(0)\bigr]+
6\bigl[(1), 4(\frac{1}{2}), 6(0)\bigr]\rightarrow\\ \bigl[(2),
3(\frac{3}{2}), 3(1), 1(\frac{1}{2})\bigr]+6\bigl[(1),
4(\frac{1}{2}),6(0)\bigr]\end{eqnarray*} In $N=3$ the long spin
3/2 multiplet is formed by adding 3 massless vector  multiplets to
the $\lambda_{MAX}=3/2$ multiplets. There remain three massless
vector multiplets. The scalar manifold of the theory is

$$\frac{\rSU(3,3)}{\rSU(3)\times\rSU(3)\times\rU(1)} $$
with $t'=9$. So we have $t-t'=15-9=6$ an $n_v-n_v'=12-6=6$.

\subparagraph{$N=4\rightarrow N'=2$.} We take $N=4$ supergravity
with 6 massless   vector multiplets. The decomposition of the
graviton multiplet  into massless multiplets of $N'=2$ is
 \begin{eqnarray*}\bigl[(2), 4(\frac{3}{2}), 6(1), 4(\frac{1}{2}),2(0)\bigr]\rightarrow
\bigl[(2), 2(\frac{3}{2}), (1)\bigr]+\\2\bigl[(\frac{3}{2}), 2(1),
(\frac{1}{2})\bigr]+\bigl[ (1), 2(\frac{1}{2}),
2(0)\bigr],\end{eqnarray*} and the decomposition of the vector
multiplet is
\begin{eqnarray*}\bigl[ (1)+4(\frac{1}{2})+6(0)\bigr]\rightarrow
\bigl[ (1)+2(\frac{1}{2})+2(0)\bigr]+\bigl[2(\frac{1}{2}),
4(0)\bigr].\end{eqnarray*} One can form two long spin 3/2
multiplets with the 2 massless $\lambda_{MAX}=3/2$  multiplets,
four massless vectors and two massless hypermultiplets. We are
left with the graviton multiplet, three massless vectors and 4
massless hypermultiplets. The coset space of this  theory is

$$\frac{\rSO(2,2)}{\rSO(2)\times \rSO(2)}\times
\frac{\rSO(4,4)}{\rSO(4)\times
\rSO(4)}\times\frac{\rSU(1,1)}{\rU(1)}, $$ with $t'=7$. $t$ and
$t'$ refer here to the translational isometries of the manifolds
SO(n,n)/SO(n)$\times$SO(n), with $t=15$ for $n=6$ and $t'=1+6$ for
$n=2,4$. So we have $t-t'=15-7=8$ and $n_v-n_v'=12-4=8$.

\subparagraph{$N=3\rightarrow N'=2$.} We start with the $N=3$
model with 3 vector multiplets as above. The decomposition of the
graviton multiplet is

\begin{eqnarray*}\bigl[(2), 3(\frac{3}{2}), 3(1), (\frac{1}{2})\bigr]\rightarrow
\bigl[(2), 2(\frac{3}{2}), (1)\bigr]+\bigl[(\frac{3}{2}), 2(1),
(\frac{1}{2})\bigr]\end{eqnarray*} and the decomposition of the
massless vector multiplet is
\begin{eqnarray*}\bigl[(1), 4(\frac{1}{2}, 6(0))\bigr]\rightarrow
\bigl[(1), 2(\frac{1}{2}), 2(0)\bigr]+\bigl[2(\frac{1}{2}),
4(0)\bigr].\end{eqnarray*} To form a long spin 3/2 multiplet we
need two massless vector multiplets and one hypermultiplet.  The
residual theory has then one vector multiplet and two
hypermultiplets. The coset is then
$$\frac{\rSU(1,1)}{\rU(1)}\times\frac{\rSU(2,2)}{\rSU(2)\times\rSU(2)\times
\rU(1) } $$ with $t'=5$. So we have $t-t'=9-5=4$ an
$n_v-n_v'=6-2=4$.

\section{Concluding remarks}

In this paper we have considered general features of the super
Higgs effect in extended supergravity,  based on the analysis of
massless and massive representations of $N$ extended supersymmetry
in four dimensions. The same analysis could be carried out for
higher dimensional theories as well. Many of these breaking
patterns find a realization in the Scherk-Schwarz mechanism of
supersymmetry breaking as well as in string compactifications in
presence of brane fluxes. The requirement of the super Higgs
effect with vanishing cosmological constant is satisfied in these
models by spontaneous breakdown of a certain number of abelian
gauge isometries, related to properties of the scalar manifolds in
the broken and unbroken phase. Many other situations can be
studied as for example the breaking of $N=4\rightarrow N'=3,2$ in
presence of an arbitrary number of matter multiplets. Also, the
more interesting case of $N\rightarrow N'=1$ has not been
considered here. In the case of $N'=2,1$ the broken phase may have
a complicated structure due to the integration of massive modes.
This is so because the reduction of supersymmetries is not as
predictive as in the cases with $N'\geq 3$.

Superstring compactifications in presence of brane fluxes appear
to offer  a general set up \cite{fp,kst} where many models of
spontaneous supersymmetry breaking can be realized and  an almost
realistic hierarchy of scales can be obtained.

\section*{Acknowledgements}

S. F. would like to thank the Dipartimento di Fisica, Politecnico
di Torino and M. A. Ll. the Theory Division at CERN  for their
kind hospitality during the  completion of this work.   Work
supported in part by the European Comunity's Human Potential
Program under contract HPRN-CT-2000-00131 Quantum Space-Time, in
which L. A.,  R. D. and M. A. Ll. are associated to Torino
University. The work of S. F. has also  been supported by the
D.O.E. grant DE-FG03-91ER40662, Task C.

\begin{table}[p]
\begin{center}
\begin{tabular} {|l|l|l|}
\cline{1-3}  $N$&massless $\lambda_{MAX}=2$  multiplet& massless
$\lambda_{MAX}=3/2$ multiplet
\\ \cline{1-3}&&\\
8&$\bigl[(2),8(\frac{3}{2}), 28(1), 56(\frac{1}{2}),
70(0)\bigr]$&none\\&&\\ 6&$\bigl[(2),6(\frac{3}{2}), 16(1),
26(\frac{1}{2}), 30(0)\bigr]$& $\bigl[(\frac{3}{2}), 6(1),
15(\frac{1}{2}), 20(0)\bigr]$\\&&\\ 5&$\bigl[(2),5(\frac{3}{2}),
10(1), 11(\frac{1}{2}), 10(0)\bigr]$& $\bigl[(\frac{3}{2}), 6(1),
15(\frac{1}{2}), 20(0)\bigr]$\\&&\\ 4&$\bigl[(2),4(\frac{3}{2}),
6(1), 4(\frac{1}{2}), 2(0)\bigr]$& $\bigl[(\frac{3}{2}), 4(1),
7(\frac{1}{2}), 8(0)\bigr]$\\&&\\ 3&$\bigl[(2),3(\frac{3}{2}),
3(1), (\frac{1}{2})\bigr]$& $\bigl[(\frac{3}{2}), 3(1),
3(\frac{1}{2}), 2(0)\bigr]$\\&&\\
2&$\bigl[(2),2(\frac{3}{2}),(1)\bigr]$& $\bigl[(\frac{3}{2}),
2(1), (\frac{1}{2})\bigr]$\\&&\\
1&$\bigl[(2),(\frac{3}{2})\bigr]$& $\bigl[(\frac{3}{2}),
(1)\bigr]$\\&&
\\\cline{1-3}
\end{tabular}
\caption{Massless $\lambda_{MAX}=2,3/2$
multiplets.}\label{hel23/2}
\end{center}
\end{table}

\begin{table}[p]
\begin{center}
\begin{tabular} {|l|l|l|}
\cline{1-3}  $N$& massless $\lambda_{MAX}=1$  multiplet &
massless $\lambda_{MAX}=1/2$  multiplet
\\ \cline{1-3}&&\\
8,6,5&none&none\\&&\\ 4&$\bigl[(1),4(\frac{1}{2}), 6(0)\bigr]$&
none\\&&\\ 3&$\bigl[(1),4(\frac{1}{2}), 6(0)\bigr]$& none\\&&\\
2&$\bigl[(1),2(\frac{1}{2}), 2(0)\bigr]$& $\bigl[2(\frac{1}{2}),
4(0)\bigr]$\\&&\\ 1&$\bigl[(1),(\frac{1}{2})\bigr]$&
$\bigl[(\frac{1}{2}), 2(0)\bigr]$\\&&\\ \cline{1-3}
\end{tabular}
\caption{Massless $\lambda_{MAX}=1,1/2$
multiplets.}\label{hel11/2}
\end{center}
\end{table}

\begin{table}[p]
\begin{center}
\begin{tabular} {|c|l|c|c|}
\cline{1-4}  $N$& massive spin 3/2 multiplet& long  &short
\\ \cline{1-4}&&&\\
8&none&&\\&&&\\
6&$2\times\bigl[(\frac{3}{2}),6(1),14(\frac{1}{2}),
14'(0)\bigr]$&no&$q=3,\,(\frac{1}{2}\mathrm{BPS})$\\&&& \\
5&$2\times\bigl[(\frac{3}{2}),6(1),14(\frac{1}{2}),
14'(0)\bigr]$&no&$q=2,\,(\frac{2}{5}\mathrm{BPS})$\\&&&\\
4&$2\times\bigl[(\frac{3}{2}),6(1),14(\frac{1}{2}),
14'(0)\bigr]$&no&$q=1,\,(\frac{1}{4}\mathrm{BPS})$\\&&&\\
&$2\times\bigl[(\frac{3}{2}),4(1),6(\frac{1}{2}),
4(0)\bigr]$&no&$q=2,\,(\frac{1}{2}\mathrm{BPS})$\\&&&\\
3&$\quad\;\;\;\bigl[(\frac{3}{2}),6(1),14(\frac{1}{2}),
14'(0)\bigr]$&yes&no\\&&&\\
&$2\times\bigl[(\frac{3}{2}),4(1),6(\frac{1}{2}),
4(0)\bigr]$&no&$q=1,\,(\frac{1}{3}\mathrm{BPS})$\\&&&\\
2&$\quad\;\;\;\bigl[(\frac{3}{2}),4(1),6(\frac{1}{2}),
4(0)\bigr]$&yes&no\\&&&\\
&$2\times\bigl[(\frac{3}{2}),2(1),(\frac{1}{2})\bigr]$&no&$q=1,\,(\frac{1}{2}
\mathrm{BPS})$\\&&&\\
1&$\quad\;\;\;\bigl[(\frac{3}{2}),2(1),(\frac{1}{2})\bigr]$&yes&no\\&&&\\
\cline{1-4}
\end{tabular}
\caption{Massive spin 3/2 multiplets.}\label{spin3/2}
\end{center}
\end{table}

\begin{table}[p]
\begin{center}
\begin{tabular} {|c|l|c|c|}
\cline{1-4}  $N$& massive spin 1 multiplet& long  &short
\\ \cline{1-4}&&&\\
8,6,5&none&&\\&&&\\ 4&$2\times\bigl[(1),4(\frac{1}{2}),
5(0)\bigr]$&no&$q=2,\,(\frac{1}{2}\mathrm{BPS})$\\&&& \\
3&$2\times\bigl[(1),4(\frac{1}{2}),
5(0)\bigr]$&no&$q=1,\,(\frac{1}{3}\mathrm{BPS})$\\&&&\\
2&$\quad\;\;\;\bigl[(1),4(\frac{1}{2}), 5(0)\bigr]$&yes&no\\&&&\\
&$2\times\bigl[(1),2(\frac{1}{2}),
(0)\bigr]$&no&$q=1,\,(\frac{1}{2}\mathrm{BPS})$\\&&&\\
1&$\quad\;\;\;\bigl[(1),2(\frac{1}{2}),
(0)\bigr]$&yes&no\\&&&\\\cline{1-4}
\end{tabular}
\caption{Massive spin 1 multiplets.}\label{spin1}

\end{center}
\end{table}

\begin{table}[p]
\begin{center}
\begin{tabular} {|c|l|c|c|}
\cline{1-4}  $N$& massive spin 1/2 multiplet& long  &short
\\ \cline{1-4}&&&\\
8,6,5,4,3&none&&\\&&&\\ 2&$2\times\bigl[(\frac{1}{2}),
2(0)\bigr]$&no&$q=1,\,(\frac{1}{2}\mathrm{BPS})$\\&&& \\
1&$\quad\;\;\;\bigl[(\frac{1}{2}),
2(0)\bigr]$&yes&no\\&&&\\\cline{1-4}
\end{tabular}
\caption{Massive spin 1/2 multiplets.}\label{spin1/2}
\end{center}
\end{table}

\begin{table}[p]
\begin{center}
\begin{tabular} {|c|l|c|c|}
\cline{1-4}  helicities& acquired masses&degeneracy&n. of physical
modes
\\ \cline{1-4}
2&0&1&2\\ $\frac{3}{2}$&$|m_i|$&2&16\\ 1&0&4&8\\ &$|m_i\pm m_j|,\;
i<j$&2&48\\ $\frac{1}{2}$&$|m_i|$&6&48\\ &$|m_i\pm m_j\pm m_k|,\;
i < j < k$&2&64\\0&0&6&6\\&$|m_i\pm m_j|\; i<j$&4&48\\&$|m_1\pm
m_2\pm m_3\pm m_4|$&2&16
\\\cline{1-4}
\end{tabular}
\caption{Mass spectrum of $N=8$ supergravity.}\label{spectrum}
\end{center}
\end{table}

\begin{table}[p]
\begin{center}
\begin{tabular} {|c|l|c|c|}
\cline{1-4}  helicities& acquired masses&degeneracy&n. of physical
modes
\\ \cline{1-4}
2&0&1&2\\ $\frac{3}{2}$&$|m_i|$&2&12\\ 1&0&4&8\\ &$|m_i\pm m_j|,\;
i<j$&2&24\\ $\frac{1}{2}$&$|m_i|$&8&48\\ &$|m_1\pm m_2\pm
m_3|$&2&4\\ 0&0&6&6\\&$|m_i\pm m_j|\; i<j$&4&24
\\\cline{1-4}
\end{tabular}
\caption{Mass spectrum of $N=6$ supergravity.}\label{spectrum6}
\end{center}
\end{table}


\begin{thebibliography}{99}


\bibitem{dwn}
B. de Wit and H. Nicolai, ``$N=8$ Supergravity.'' {\it Nucl.
Phys. B} {\bf 208}, 323 (1982).


\bibitem{war}
 N.P. Warner,``Some Properties of the Scalar Potential in
Gauged Supergravity Theories'',  {\it  Nucl. Phys. B} {\bf 231} 250 (1984) 



\bibitem{hw1} C. M. Hull and  N. P. Warner,
``The Potentials of the Gauged N=8 Supergravity Theories'', {\it
Nucl.  Phys. B} {\bf 253}, 675 (1985).

\bibitem{fsz} S. Ferrara, C. A. Savoy and B. Zumino, ``General Massive Multiplets
in Extended Supersymmetry''. {\it Phys. Lett. B} {\bf 100} n.5 393
(1981).

\bibitem{st} J. Strathdee, ``Extended Poincare Supersymmetry".
 {\it Int. J. Mod. Phys. A} {\bf 2} (1) 273 (1987).

\bibitem{ccfdfm} L. Castellani, A. Ceresole, S. Ferrara, R. D'Auria, P. Fre, E.
Maina, ``The Complete $N=3$ Matter Coupled Supergravity''. {\it
Nucl.Phys.B} {\bf 268} 317 (1986).


\bibitem{cgp}
S. Cecotti, L. Girardello and M. Porrati, ``An Exceptional N=2
Supergravity with Flat Potential and Partial Superhiggs''. {\it
Phys. Lett. B} {\bf 168}, 83 (1986).



\bibitem{fgp}
S. Ferrara, L. Girardello and M. Porrati, ``Minimal Higgs Branch
for the Breaking of Half of the Supersymmetries in N=2
Supergravity''. {\it Phys. Lett. B} {\bf 366}  155 (1996).



\bibitem{tz}
V. A. Tsokur and Y. M. Zinovev, ``Spontaneous Supersymmetry
Breaking in N = 4 Supergravity with Matter,'' {\it Phys.  Atom.
Nucl.}  {\bf 59}, 2192 (1996);``Spontaneous Supersymmetry Breaking
in N = 3 Supergravity with Matter''. {\it Phys.  Atom.  Nucl.}
{\bf 59} 2185 (1996).



\bibitem{ss}
J. Scherk and J. H. Schwarz, ``How To Get Masses From Extra
Dimensions''. {\it Nucl. Phys. B} {\bf 153}, 61 (1979).

\bibitem{css} E. Cremmer, J. Scherk and J. H. Schwarz,
 ``Spontaneously Broken N=8 Supergravity''.
{\it Phys. Lett. B} {\bf 84}, 83 (1979).



\bibitem{fp}
A. R. Frey and J. Polchinski, ``N = 3 warped compactifications''.
hep-th/0201029.



\bibitem{kst}
S. Kachru, M. Schulz and S. Trivedi,
 ``Moduli Stabilization from Fluxes in a Simple IIB Orientifold''.
 hep-th/0201028.

\bibitem{ps}
J. Polchinski and A. Strominger, ``New Vacua for Type II String
Theory''. {\it Phys.  Lett.  B} {\bf 388}, 736 (1996).



\bibitem{tv}
T. R. Taylor and C. Vafa, ``RR flux on Calabi-Yau and partial
supersymmetry breaking''. {\it Phys. Lett. B} {\bf 474}, 130
(2000).

\bibitem{ma}
P. Mayr, ``On Supersymmetry Breaking in String Theory and its
Realization in Brane Worlds''. {\it Nucl.  Phys.  B} {\bf 593}, 99
(2001).

\bibitem{cklt}
G. Curio, A. Klemm, D. Lust and S. Theisen, ``On the vacuum
structure of type II string compactifications on  Calabi-Yau
spaces with H-fluxes''. {\it Nucl. Phys. B} {\bf 609}, 3 (2001).


\bibitem{gkp} S. B. Giddings, S. Kachru and J. Polchinski,
``Hierarchies from Fluxes in String Compactifications''.
hep-th/0105097.


\bibitem{hu}
C.M. Hull, `` Noncompact Gaugings of $N=8$ Supergravity.'' {\it
Phys. Lett. B} {\bf 142}, 39 (1984); `` More Gaugings of $N=8$
Supergravity.'' {\it Phys. Lett. B} {\bf 148}, 297 (1984).


\bibitem{hw}
C. M. Hull and N. P. Warner, ``The Structure Of The Gauged N=8
Supergravity Theories''. {\it Nucl.  Phys. B} {\bf 253}, 650
(1985).




\bibitem{cfgtt}
F. Cordaro, P. Fre, L. Gualtieri, P. Termonia and M. Trigiante,
``N = 8 Gaugings Revisited: An Exhaustive Classification''. {\it
Nucl. Phys. B } {\bf 532}, 245 (1998.)

\bibitem{dwn2}
B.~de Wit and H.~Nicolai,
``The Consistency Of The S**7 Truncation In D = 11 Supergravity,''
{\it Nucl.\ Phys.\ B} {\bf 281} (1987) 211.

\bibitem{dlp}

M. J. Duff and C. N. Pope, ``Consistent Truncations in
Kaluza-Klein Theories''. {\it Nucl. Phys. B }{\bf 255} (1985) 355.

M. J. Duff, B. E. Nilsson, N. P. Warner and C. N. Pope,``On The
Consistency Of The Kaluza-Klein Ansatz'', {\it Phys. Lett. B} {\bf
149}, 90 (1984).


M. J. Duff, H. Lu and C. N. Pope, ``AdS(5) x S(5) Untwisted''.
{\it Nucl. Phys. B} {\bf 532} 181  (1998).

M.~Cvetic, H. Lu and C. N. Pope, ``Consistent Kaluza-Klein sphere
reductions'', {\it Phys. Rev. D} {\bf 62}, 064028 (2000).


M. Cvetic, H. Lu, C. N. Pope, A. Sadrzadeh and T. A. Tran,
``Consistent SO(6) reduction of type IIB supergravity on S(5)'',
{\it Nucl.  Phys.  B}  {\bf 586}, 275 (2000).

\bibitem{adfl} L. Andrianopoli, R. D'Auria, S. Ferrara and M. A. Lled\'o,
``Gauging of Flat Groups in Four Dimensional Supergravity'',
hep-th/0203206; ``Duality and Spontaneously Broken Supergravity in
Flat Backgrounds'', hep-th/0204145.




\bibitem{bkl}
R. Blumenhagen, C. Kounnas and D. Lust, ``Continuous Gauge and
Supersymmetry Breaking for Open Strings on  D-branes''. {\it JHEP}
{\bf 0001}, 036 (2000);

I. Antoniadis, J. P. Derendinger and C. Kounnas,
``Non-perturbative Supersymmetry Breaking and Finite Temperature
Instabilities in N = 4 Superstrings''. hep-th/9908137.


\bibitem{kk} E. Kiritsis and C. Kounnas, ``Perturbative and Non-perturbative
Partial Supersymmetry Breaking:  N = 4 $\to$ N = 2 $\to$ N = 1''.
{\it Nucl. Phys. B }{\bf 503} 117 (1997).






\bibitem{dfl} R. D'Auria, S. Ferrara and M. A. Lled\'o,
``On Central Charges and Hamiltonians for O-brane dynamics''. {\it Phys. Rev. D} {\bf 60}
 084007 (1999).


\bibitem{fz} S. Ferrara, B. Zumino, ``The Mass Matrix of $N=8$ Supergravity''. {\it Phys. Lett. B} {\bf 86} ns. 3,4 279 (1979).


\bibitem{he} S. Helgason, ``Differential Geometry, Lie Groups and Symmetric Spaces''. Academic Press, (1978).

\bibitem{adfft}
L. Andrianopoli, R. D'Auria, S. Ferrara, P. Fr\'e and M.
Trigiante, ``R-R scalars, U-duality and Solvable Lie Algebras''.
{\it Nucl.  Phys.  B}  {\bf 496} 617 (1997);

L. Andrianopoli, R. D'Auria, S. Ferrara, P. Fr\'e, R. Minasian and
M. Trigiante, ``Solvable Lie algebras in Type IIA, Type IIB and M
Theories''. {\it Nucl. Phys. B} {\bf 493} 249 (1997).

\bibitem{lps}
H. Lu, C. N. Pope and K. S. Stelle, ``Weyl Group Invariance and
p-brane Multiplets''. {\it Nucl.  Phys. B} {\bf 476} 89 (1996).


\bibitem{cjlp}
E. Cremmer, B. Julia, H. Lu and C. N. Pope, ``Dualisation of
Dualities. II: Twisted Self-Duality of Doubled Fields  and
Superdualities''. {\it Nucl. Phys.  B} {\bf 535} 242 (1998);

E. Cremmer, B. Julia, H. Lu and C. N. Pope, ``Higher-Dimensional
Origin of D = 3 Coset Symmetries''. hep-th/9909099.



\bibitem{svn}
E. Sezgin and P. van Nieuwenhuizen, ``Renormalizability Properties
of Spontaneously Broken N=8 Supergravity''. {\it Nucl. Phys.  B}
{\bf 195}, 325 (1982).


\bibitem{adf}
L. Andrianopoli, R. D'Auria and S. Ferrara, ``Supersymmetry
Reduction of N-extended Supergravities in Four  Dimensions''.
hep-th/0110277.


\end{thebibliography}
\end{document}